\documentclass[a4paper,12pt]{article}
\usepackage[english]{babel}
\usepackage[utf8]{inputenc}	
\usepackage[T1]{fontenc}
\usepackage{times}
\usepackage{amsfonts,amsmath,amssymb,amsthm}
\usepackage{bm} 		
\usepackage{graphics,graphicx}			
\usepackage{multicol}
\usepackage{float}
\usepackage{subfig}
\usepackage{lineno}
\usepackage{setspace}

\date{}	
\newcommand{\pt}{\partial}

\theoremstyle{definition}

\begin{document}
\title{\textbf{\fontsize{12}{12}A sewage management proposal for Luruaco lake, Colombia}} 
\author{
 T. M. SAITA \\
Departamento de Matemática \\
 Universidade Tecnológica Federal do Paraná - Cornélio Procópio - PR - Brasil \\
 tatianasaita@utfpr.edu.br \\
 \\
 P. L. NATTI, E.R.CIRILO, N.M.L. ROMEIRO \\ 
Departamento de Matemática \\ 	
Universidade Estadual de Londrina - Londrina - PR - Brasil \\
	plnatti@uel.br, ercirilo@uel.br, nromeiro@uel.br  \\
   \\
M.A.C. CANDEZANO \\
Centro de Modelación Matemática y Computación Científica \\
Universidad del Atlántico - Barranquilla - Atlantico -  Colombia \\
miguelcaro@mail.uniatlantico.edu.co \\
\\
R.B. ACUÑA, L.C.G. MORENO \\
Departamento de Biología  \\ 
Universidad del Atlántico - Barranquilla - Atlántico - Colombia \\
rafaelborja@mail.uniatlantico.edu.co, luiscarlosgutierrez@mail.uniatlantico.edu.co}
	
\maketitle

\begin{abstract}
	
	This study presents numerical simulations of faecal coliforms dynamics in Luruaco lake, located in Atlántico Department, Colombia. The velocity field is obtained through a two-dimensional horizontal (2DH) model of Navier-Stokes equations system. The transport equation of faecal coliforms concentration is provided from a convective-diffusive-reactive equation. The lake’s geometry is built through cubic spline and multiblock methods. The discretization method by Finite Differences and the First Order Upwind (FOU) are applied to the 2DH model. The Mark and Cell (MAC) method is used to determine numerically the velocity field of water flow.  Numerical simulations are carried out for a 72-hour period in order to understand the influence of faecal coliforms injections from each tributary. From the qualitative and quantitative analysis of the factors that influence faecal coliforms dynamics,   proposals are presented, which aims to reduce contamination in some regions of Lake Luruaco. The numerical simulations show that the best option to improve water quality in lake is the implementation of two actions, the diversion of the Limón stream to the Negro stream and the installation of a sewage treatment plant at the mouth of the Negro stream. Other less expensive proposals are also presented.

\end{abstract}

\section{Introduction}

Water pollution is a global problem. Factors as the use of pesticides, the deforestation, the forest burning, and the industrial and domestic sewage can contribute to pollution in water bodies, by altering physicochemical characteristics of water and causing health risks to those using these contaminated waters (Helmer and Hespanhol, 1997; Burneo and Gunkel, 2003; Pardo et al., 2012; Jerves-Cobo et al., 2018; Wang et al., 2019). The World Health Organization (WHO) regulates the limits of some physicochemical parameters, such as pH, temperature, presence of substances like aluminium and nitrite, so that water can be safely destined human consumption. These water quality parameters serve as a tool for government agencies to maintain water resources and develop strategies for the management of water bodies (OPS, 1988; WHO, 2017). 

The Luruaco lake is an example of water body that is affected by pollution. Located in the Atlántico Department, in Colombia, it is used as water source to supply approximately 28,000 local residents (Municipio de Luruaco, 2012). The lake is contaminated by organic matter waste, generated mainly by domestic sewage from Luruaco population, which gets in the lake through its tributaries (CRA, 2012). Domestic sewage contain pathogenic microorganisms such as bacteria, viruses or protozoa, that cause diseases like diarrhea and others infections, especially in children. Therefore, the continuous analysis of the water quality of Luruaco lake is an effective way to assess the risks of contamination of the population using this water (CRA, 2014). 

The WHO uses faecal coliform bacteria as indicator of the presence of pathogenic organisms (DHSS, 1982; WHO, 2017). Thus, local monitoring of the concentration of faecal coliforms is a method to identify the intensity of contamination in lakes (Reckhow and Chapra, 1983; Janssen and Heuberger, 1995; Brauwere et al., 2014; VishnuRadhan et al., 2018; Ouattara et al., 2018). In this way, based on experimental measurements, mathematical models can be calibrated to simulate the dynamics of faecal coliforms in lakes, allowing government agencies to create strategies, for example, to capture water from lake regions with less contamination by faecal coliform. 

In this work the numerical simulations consider the flow in Lake Luruaco as laminar, since its average depth is 5 meters, while its area is $2.5 \times 10^6 m^2 $ (CRA, 2014). These characteristics allow to model the water flow at lake by means of a laminar two-dimensional horizontal (2DH) model (Horita and Rosman, 2006). The mathematical model in generalized coordinates (Maliska and Raithby, 1984; Maliska, 2004) describes the faecal coliform dynamics on the lake’s surface, taking into account the velocity fields obtained from hydrodynamic model formed by the mass conservation and the Navier-Stokes equations (Pardo et al., 2012; Romeiro et al., 2017). About the transport model, it is considered that the transport of faecal coliforms in the lake is passive, i.e, the coliforms do not significantly modify the water flow, allowing the velocity field (obtained through a hydrodynamic model) to be calculated separately from the model of faecal coliforms concentration. The transport model is given by a convective-diffusive-reactive equation (Romeiro et al., 2011; Saita et al., 2017).

In order to have the analytical mathematical model simulated computationally, the discretization of equations is carried out through the finite difference method (Evans et al., 1999; Fortuna, 2012). Given the dynamic complexity generated by the nonlinear terms of the hydrodynamic model, we use the First Order Upwind method and the Mark and Cell method (MAC) to solve the hydrodynamic model (Mckee et al., 2008; Cirilo et al., 2018). Thus we obtain the velocity fields of water flow, necessary to simulate the transport model of faecal coliforms concentration.

The construction of Luruaco lake geometry takes into considerations its real characteristics. The irregular contour is described by cubic splines, while the interior of the lake is described by the multiblock method. These procedures create a computational grid that adjusts best to the type of geometry without requiring complex computer calculations (Pardo et al., 2012; Saita et al., 2017; Romeiro et al., 2017). Given real initial and boundary conditions, the simulations point to the influence of tributaries on faecal coliform dynamics. After 72 hours of numerical simulation, the dynamics of faecal coliforms reach stationary characteristics.

The results obtained in this study allowed us to understand how some factors influence the dynamics of faecal coliforms in the lake, among which we highlight the influence of tributaries and the existence of hydrodynamic vortices. Thus, from the numerical simulations, it was possible to elaborate a management proposal for Luruaco lake. 

Obviously, due to the costs, the installation of treatment plants in all tributaries of Lake Luruaco is not feasible. Even the installation of a single treatment plant is a challenge. In this context, we suggest that the Limón stream is diverted to the Negro stream and that a single sewage treatment plant be installed at the mouth of the  Negro stream. Numerical simulations show that our proposal creates large regions in Luruaco lake with very low faecal coliforms pollution. These new regions with little contamination by microorganisms could be new water collection points for the region's residents. Due to the characteristics of the hydrodynamic flow of Luruaco Lake, we emphasize that the installation of single sewage treatment plants, in the other tributaries, does not generate significant improvements in the water quality of Luruaco Lake. Other less expensive proposals are also presented.

\section{The Luruaco Lake}

Luruaco lake located in the Department of Atlántico, Colombia, more precisely between the coordinates 10º 16' and 11º 04' north latitude and 74º 43' and 75º 16' west longitude, it occupies an area of 420 hectares. The lake has an average depth of 5 meters, with a volume of approximately 12,5 $\times $10$^6$ m$^3$ (CRA, 2014). Situated at 31 m of altitude, the region's climate is  tropical, with temperatures between 24ºC and 28ºC, prevailing the hot and dry climate during the year (CRA, 2012). 

The Luruaco lake belongs to the basin of the Canal del Dique and is supplied by smaller tributaries (CRA,2007). The main ones are the Mateo stream, Negro stream and Limón stream (CRA, 2012). The canal that connects Luruaco lake to San Juan del Tocágua lake, due to the difference of altitude, serves as effluent of the Luruaco lake, as shown in the figure 1.

\begin{figure}[!h]
	\begin{center}
		\includegraphics[scale=0.9]{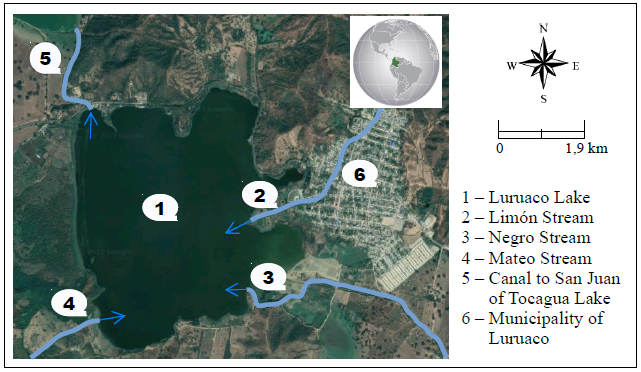} 
		\caption{Localization of Luruaco lake and tributaries (Limón, Negro, Mateo). Font: Google Maps/Author.}
		\label{fig1}
	\end{center}
\end{figure}

Close to the east shore of the lake is situated the municipality of Luruaco with about 28 000 inhabitants. The residents of the municipality use the lake for fishing and tourism, they also use their water for agriculture and livestock, but the main utility of the lake is the supply of water for consumption by the population. (MUNICIPIO DE LURUACO, 2012). 

The lake suffers from pollution, mainly from domestic sewage generated by the municipality of Luruaco, which is injected into the lake without proper treatment (CRA, 2012). Consumption of water contaminated with pathogenic microorganisms can lead to the emergence of various diseases. In this sense, it is important to analyze the dynamics and concentrations of faecal coliforms in Lake Luruaco, since such concentrations are used as a biological parameter in the evaluation of water quality. (OPS, 1988; Ashbolt et al., 2001).

In this work, the numerical simulations of faecal coliform dynamics aimed to indicate the regions of Lake Luruaco with the highest and lowest  concentrations of faecal coliforms. We emphasize that to characterize the problem under study, we must use mathematical and numerical models calibrated with the physical properties and local environmental conditions, among other factors. 

\section{Development of mathematical and numerical model}
\label{sec2}

About the irregular geometry of the lake, so that the characteristics of the lake geometry are present in the model, the computational grid (domain where the model will be simulated) is constructed in generalized coordinates $(\xi, \eta)$ (Maliska and Raithby, 1984; Maliska, 2004; Pardo et al., 2012; Saita et al., 2017). The coordinates of the points on the lake border are collected using Webplottdigitizer software from Google Maps photos (Rohatgi, 2018). Later, the points are interpolated through the cubic spline method. The grid is constructed using the multiblock method. These procedures create a computational grid that adjusts best to the type of geometry without requiring complex computer calculations (Pardo et al., 2012; Saita et al., 2017; Romeiro et al., 2017).

The water flow of Igapó I Lake is laminar (low declivity and the small volume of input water compared with water volume of the Lake), so that a two-dimensional horizontal (2DH) hydrodynamic model represents it appropriately (HORITA; ROSMAN, 2006). Considering that the fluid (water) is incompressible, of the Newtonian type and in hydrostatic equilibrium, so the variations of density are not significant 
(SHAKIB et al., 1991; SCHLICHTING; GERTEN, 2000). It is assumed that the fluid (water) has constant viscosity. It is also considered that the transport of faecal coliforms in the lake is passive, i.e, the coliforms do not significantly modify the water flow, allowing the velocity field (obtained through a hydrodynamic model) to be calculated separately from the model of faecal coliforms concentration. Assuming as well that the forces of the external field (actions of wind, heat, ...) are not expressive and that there are no variations in the Lake’s boundary, the water velocity field  is obtained by means of a hydrodynamic model given by the equation for mass conservation \eqref{1} and the Navier-Stokes equations \eqref{2} - \eqref{3} in generalized coordinates $(\xi, \eta, \tau)$ bellow: 

 \begin{eqnarray} 
 \underbrace{\frac{\pt U}{\pt \xi}+ \frac{\pt V}{ \pt \eta}}_{\mbox{{\small mass conservation term}}} &=&0 \;, \label{1} \\
 \nonumber \\
 \underbrace{\frac{\pt}{\pt \tau} \left(\frac{ u}{J} \right)}_{\mbox{{\small temporal term}}} +\underbrace{\frac{\pt}{\pt \xi} ( U u ) +\frac{\pt}{\pt \eta} ( V u)}_{\mbox{{\small convective term}}} & = & \frac{1}{\rho} \underbrace{\left(\frac{\pt p}{\pt \eta} \frac{\pt y}{\pt \xi}-\frac{\pt p}{\pt \xi} \frac{\pt y}{\pt \eta} \right)}_{\mbox{{\small pressure term}}} + \nu \underbrace{\left[\frac{\pt}{\pt \xi} \left( J \left( D \frac{\pt u}{\pt \xi} -D \frac{\pt u}{\pt \eta} \right)\right)\right.}_{\mbox{{\small diffusive term}}}  \nonumber \\
 & + & \underbrace{ \left. \frac{\pt }{\pt \eta} \left(J \left( D \frac{\pt u}{\pt \eta} - D \frac{\pt u}{\pt \xi} \right) \right) \right]}_{\mbox{{\small diffusive term}}}, \label{2}\\
 \nonumber \\
 \underbrace{\frac{\pt}{\pt \tau} \left(\frac{ v}{J} \right)}_{\mbox{{\small temporal term}}} + \underbrace{\frac{\pt}{\pt \xi} ( U v ) +\frac{\pt}{\pt \eta} ( V v)}_{\mbox{{\small convective term}}} & = & \underbrace{\frac{1}{J} \left(\frac{\pt p}{\pt \xi} \frac{\pt x}{\pt \eta} -\frac{\pt p}{\pt \eta} \frac{\pt x}{\pt \xi} \right)}_{\mbox{{\small pressure term}}} +\nu \underbrace{\left[\frac{\pt}{\pt \xi} \left( J \left( D \frac{\pt v}{\pt \xi} -D \frac{\pt v}{\pt \eta} \right)\right) \right.}_{\mbox{{\small diffusive term}}} \nonumber \\
 & + &  \underbrace{\left. \frac{\pt }{\pt \eta} \left(J \left( D \frac{\pt v}{\pt \eta} - D \frac{\pt v}{\pt \xi} \right) \right) \right]}_{\mbox{{\small diffusive term}}},
 \label{3}
 \end{eqnarray}
where $u$ and $v$ are the components of the velocity vector of water flow, $p$ is the pressure, $\rho$ and $\nu$ are constants representing the density and viscosity of the water, respectively. The terms $U$ and $V$ are the components of contravariant velocities, $J$ is the Jacobian of the transformation of cartesian coordinates $x$ and $y$ to the generalized coordinates $\xi$ and $\eta$, and $D$ is the diffusion coefficient of faecal coliforms, equal in the directions $\xi$ and $\eta$ (Pardo et al., 2012; Saita et al., 2017; Romeiro et al., 2017).

With the water velocity field obtained from the system (1)-(3), it is possible to find the concentration of faecal coliforms given by the transport model, a convective-diffusive-reactive PDE,
 \begin{eqnarray}
 \underbrace{\frac{\pt}{\pt \tau} \left(\frac{C}{J} \right)}_{\mbox{{\small  temporal term}}}+\underbrace{\frac{\pt}{\pt \xi} (UC ) +\frac{\pt}{\pt \eta} (VC)}_{\mbox{{\small convective term}}}
 & = &  - \underbrace{\frac{KC}{J}}_{\mbox{{\small reactive term}}} + \underbrace{\frac{\pt}{\pt \xi} \left[ J \left(D \frac{\pt C}{\pt \xi} - D \frac{\pt C}{\pt \eta} \right) \right.}_{\mbox{{\small diffusive term}}}
 + \nonumber \\
 \nonumber \\
 & + &  \underbrace{ \frac{\pt }{\pt \eta} \left. J \left( D \frac{\pt C}{\pt \eta} - D \frac{\pt C}{\pt \xi} \right) \right],}_{\mbox{{\small diffusive term}}} 
 \label{4}
 \end{eqnarray}
where $C=C(\xi, \eta, \tau)$ represents the concentration of faecal coliforms and $K$ is the decay constant (Pardo et al., 2012; Saita et al., 2017; Romeiro et al., 2017).

Note that the terms of equations (1)-(4) are grouped according to the characteristics: pressure terms, temporal terms, convective terms, diffusive terms and reactive terms. In numerical model, the equations (1)-(4) are discretized by means of the finite differences method (FDM). The term convective has nonlinearity, so we use the First Order Upwind (FOU) method that provides consistent numerical results for the model (Cirilo et al., 2018). 
 
\section{Initial and  boundary conditions}
\label{sec3}

Next, initial and boundary conditions are imposed so that the numerical model, developed in Section 3, characterizes the dynamics of faecal coliforms in Lake Luruaco. The continuous injection of water with faecal coliform through the Mateo, Negro and Limón streams is considered. The drainage of the Luruaco lake is made through the canal that connects it to the San Juan del Tocágua lake (see figure 1).

On the hydrodynamic model (1)-(3), it is considered that the water flow has Reynolds number, $Re = 555$, a value that characterizes the flow of water in the Luruaco lake.

About the molecular diffusion coefficient $D$ and decay coefficient $K$ of faecal coliforms, they are considered constant throughout the geometry of the lake. The molecular diffusion coefficient $D$ locally spreads faecal coliforms by random thermal movement. However, on a large scale, it is vortices and swirls that spread the faecal coliforms through the so-called turbulent diffusion. The magnitude order of the turbulent diffusion is greater than the order of magnitude of the molecular diffusion, since the scale of the vortex is much larger than the scale of molecular diffusion. According to the Chapra (2008), the various reactive species have molecular diffusion values in the interval between $D = 10^{-3} m^2/h$ and $D = 10^{-1} m^2/h$. The turbulent diffusion coefficient in lakes, which varies with the scale of the turbulent phenomenon, assumes values between $D = 10^1 m^2/h$ and $D = 10^{10} m^2/h$. In order to simulate the molecular and turbulent  diffusion phenomena involved in faecal coliforms transport in Luruaco lake, the best fit for the diffusion coefficient is given by $D = 3.6 m^2/h$. For decay coefficient of faecal coliforms, the best fit is $K =0.02/h$ (Romeiro et al., 2011; Liu et al., 2015; Romeiro et al., 2017). 

In addition, the initial and boundary conditions for the numerical model are considered:
\begin{itemize}
	\item \textbf{Initial condition for the hydrodynamic model:} We performed simulations starting from a velocity and pressure fields in a state of quiescence. We understand the quiescence state as the one with null velocity and pressure fields.
	\item \textbf{Boundary condition for the hydrodynamic model:} For $t>0$ it is considered that velocity of water flow is null throughout its boundary, except in the entrances and exit of the lake. In entrances and exit of the lake is considered Neumann’s condition boundary (FORTUNA, 2012; BURDEN et al., 2015). In relation to the pressure field, it is considered a gradient of 10 $\%$ between entrance pressure and exit pressure of Luruaco Lake, while in the other points of the boundary the Neumann’s boundary condition is used.
	\item \textbf{Initial condition for the faecal coliform concentration model:} At the instant $t=0$ the faecal coliform concentration field is null in the interior and at the boundary of the lake.
	\item \textbf{Boundary condition for faecal coliform concentration model:} For $t>0$ it is considered that in the entrances of Luruaco lake  there are continuous injections of faecal coliforms, with constant concentrations, over the 72-hour period to achieve steady state. For each simulation, the faecal coliforms concentration values will be indicated.
 \end{itemize}

In this context, the simulation of the numerical model describes qualitatively the impact that a continuous discharge of faecal coliforms, in the 3 entrances (streams) of Luruaco lake, produces in all its extension.

\section{Numerical simulations}
\label{sec4}

In this section, the dynamics of faecal coliforms will be qualitatively analyzed. It is important to highlight that the objective of these simulations is not to offer accurate values of faecal coliform concentration, but we want to present the influence of each factor considered in the modeling (water velocity field, streams and lake geometry) on the faecal coliform dynamics in the lake. 

The computational code is developed in Fortran 90 language. The hydrodynamic model is calculated using Mark and Cell method (Mckee et al., 2008; Cirilo et al., 2018), generating the velocity field of water flow. Simulation of faecal coliform injection through the three entrances (streams) is then performed separately, which allowed us to analyze the impact of each stream on lake pollution. Finally, we present the simulation with the injection of faecal coliforms by the three tributaries simultaneously. From these results it was possible to develop a management proposal for the discharge of sewage in the lake, which possibly will allow the appearance of regions located in the lake with better water quality.

\subsection{Velocity field of water flow}
\label{sec5}
\begin{figure}[H]
	\begin{center}
		\includegraphics[scale=0.9]{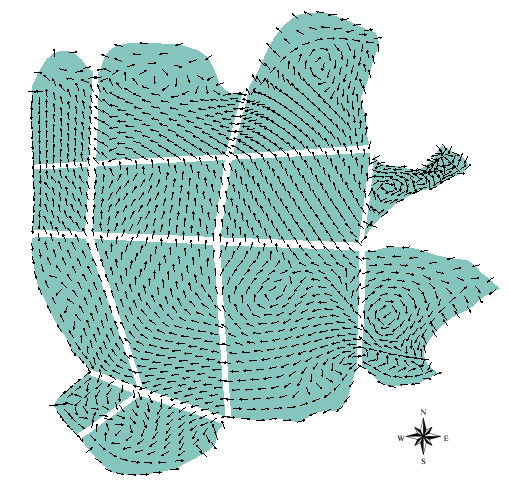} 
		\caption{Steady water flow in Luruaco Lake. Font: Author.}
		\label{fig2}
	\end{center}
\end{figure}

We performed the simulation over a 72 hour period to achieve steady state. Under the conditions presented in Section \ref{sec3}, the steady water flow is obtained, as can be seen in Figure \ref{fig2}.

Note that due to the application of the multiblock method for the construction of the computational grid, in this study the lake was constructed by 13 blocks, with boundary conditions between the blocks that provide the representation of the flow in the total lake area during the simulation (Saita et al., 2017).

Furthermore, it is possible to observe the formation of vortices in several regions of the lake, increasing the flow rate in some points. On the other hand, as in the "arm" to the east of the lake, the vortex formed at its entrance does not allow the passage of water to this location, so that this region has little water renewal.



\subsection{Contamination by Mateo stream}
\label{sec6}
\begin{figure}[H]
 	\begin{center}
 		\includegraphics[scale=0.8]{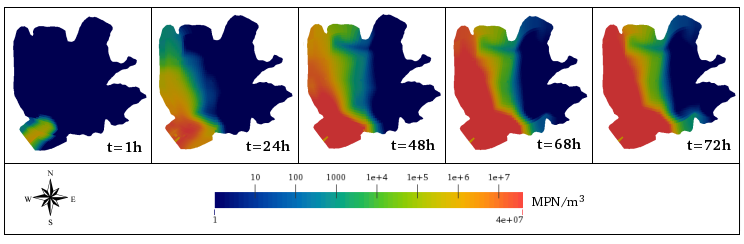} 
 		\caption{Simulation of faecal coliform injection in Luruaco lake through Mateo stream. Font: Author.}
 		\label{fig3}
 	\end{center}
 \end{figure}

In this simulation, the water injection is continuous through the three tributaries (streams), maintaining the same velocity field analyzed in the Section \ref{sec5}. On the other hand, the faecal coliforms are injected only through the Mateo stream located to the southwest of the lake. In this simulation $C_1= 100 MPN/100 ml$ is the faecal coliforms concentration injected. In Figure \ref{fig3} we can see the dynamics of faecal coliform contamination due to the Mateo stream.

Due to the velocity field (see Figure \ref{fig2}) and the location of Mateo stream, faecal coliforms tend to focus on the west region of the lake, then being drained out of Luruaco lake by the canal to San Juan of Tocagua lake (see Figure 1).

Another factor influencing the dynamics of faecal coliforms is the decay property, that is, the rate at which faecal coliforms degrade over time through natural processes (Chapra, 2008; Romeiro et al., 2011). This characteristic is being considered in the mathematical model, through the term $K=0.02/h$ in the transport model of faecal coliforms (see equation \eqref{4}). The decay also explains the fact that faecal coliforms do not spread throughout the lake.
 
Therefore, considering only the concentration of faecal coliforms $C_1$ injected by the Mateo stream, it is noticeable that the region near the municipality of Luruaco, where there is more water catchment for human consumption, does not present high levels of concentration of faecal coliforms due to this affluent.
 
 \subsection{Contamination by Negro stream}
 \label{sec7}
\begin{figure}[H]
 	\begin{center}
 		\includegraphics[scale=0.8]{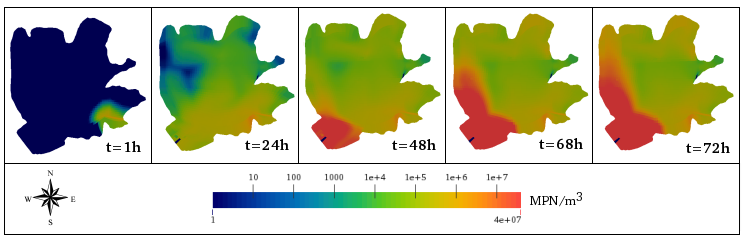} 
 		\caption{Simulation of faecal coliform injection in Luruaco lake through Negro stream. Font: Author.}
 		\label{fig4}
 	\end{center}
 \end{figure}

In this simulation, the water injection is continuous through the three tributaries (streams), maintaining the same velocity field analyzed in the Section \ref{sec5}. On the other hand, the faecal coliforms are injected only through the Negro stream located to the southeast of Luruaco lake. In this simulation $C_2= 100 MPN/100 ml$ is the faecal coliforms concentration injected. In Figure \ref{fig4} we can see the dynamics of faecal coliform contamination due to the Negro stream.
 
In the simulation, it is observed that, contrary to the pattern observed for contamination due to the Mateo stream, the faecal coliforms spread over the entire length of the lake. This dynamics occurs due to the influence of vortices in the velocity field that intensify the spread of faecal coliforms. The existence of a vortex near the entrance of the Negro stream results in the faecal coliforms not to concentrate in this region, but end up flowing mainly to the region close to the Mateo River and then to the exit of the Luruaco lake. 
The Figure \ref{fig4} shows a high concentration of faecal coliforms in the southwest of the lake.

\subsection{Contamination by Limón stream}
 \label{sec8}
 
\begin{figure}[H]
  	\begin{center}
  		\includegraphics[scale=0.8]{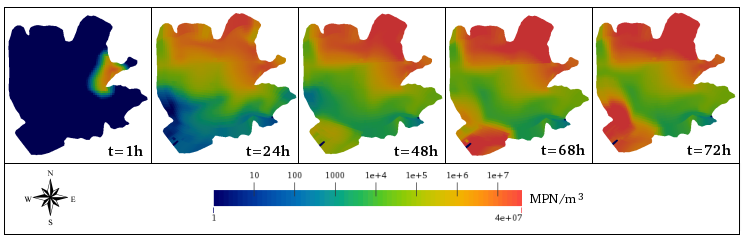} 
  		\caption{Simulation of faecal coliform injection in Luruaco lake through Limón stream. Font: Author.}
  		\label{fig5}
  	\end{center}
  \end{figure}

The Limón stream runs through the municipality of Luruaco. As the municipality does not have a basic sanitation system for the entire population, the Limón stream receives more sewage than the Negro and Mateo streams (CRA, 2012). In this simulation $C_3= 500 MPN/100 ml$ is the faecal coliforms concentration injected, a concentration higher than that used in previous simulations. In Figure \ref{fig5} we can see the dynamics of faecal coliform contamination due to the Limón stream.
 
Again, it can be seen the influence of the field of velocity in the dynamics. Vortices spread faecal coliforms throughout the lake, especially to the north and southwest regions of the lake. Note that the southeast region of the lake is the least affected region due to the injection of faecal coliforms by the Limon stream.

The lake region near the mouth of the Limón stream is the main source of water for residents of the municipality of Luruaco. However, the simulation shows that the concentration of faecal coliforms in this region is high, increasing the risk of contamination of the population.  
  
  \subsection{Contamination by Mateo, Negro and Limón streams}
	\label{sec8a}
	
\begin{figure}[H]
   	\begin{center}
   		\includegraphics[scale=0.8]{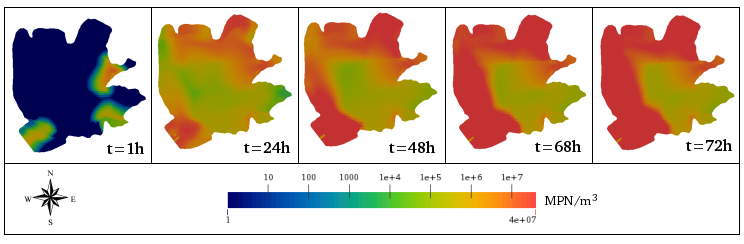} 
   		\caption{Simulation of faecal coliforms injection in Luruaco lake through the Mateo, Negro and Limón streams, simultaneously. Font: Author.}
   		\label{fig6}
   	\end{center}
   \end{figure}

The previous dynamics of faecal coliform concentrations analyzed the influence of each stream in Luruaco lake contamination, separately. The simulations showed different behaviors for each stream. The following simulation aims to analyze the real dynamics of faecal coliforms in the lake, when there is injection through the three streams, Mateo, Negro and Limón, simultaneously. Consider the following faecal coliform concentrations: 
$C_1=100 MPN/100 ml$ by Mateo stream, $C_2=100 MPN/100 ml$ by Negro stream and $C_3=500 MPN/100 ml$ by Limón stream. Figure \ref{fig6} shows the faecal coliform concentrations in Luruaco lake due to Mateo, Negro and Limón streams.
  
	Note that the characteristics of the simulations seen in sections \ref{sec6} - \ref{sec8} are also present in this simulation. Figure \ref{fig6} shows that the central and southeast regions have the lowest concentrations of faecal coliforms, indicating that in the current scenario these regions can provide water with less risk to the health of the population. This result is the basis of our proposal for sewage management to improve the water quality that supplies the municipality of Luruaco.
   
 \section{Proposals for sewage management in tributaries}
 \label{sec9}
 
 Through simulations, it is possible to qualitatively analyze the dynamics and characteristics that influence faecal coliform concentrations in Luruaco Lake. It can be seen that the geometry of the lake, the streams and vortices, and the conditions imposed on the model influence the simulations. For example, even if we inject a low concentration of faecal coliforms through the Negro stream, the vortices tends to spread it, and in the stationary situation, there will be high concentration of faecal coliforms in the southwest region of the lake. This is a pattern that occurs in Figures 4 and 6, a consequence of the vortices in the southeast and central regions of the lake.

We would also like to comment that, due to the costs, the installation of sewage treatment plants in all tributaries of Luruaco lake is not feasible. Even the installation of a single sewage treatment plant is a challenge. 

Below we present our proposals. We emphasize that economic, technical, topological and environmental viability studies, among others, must be carried out to verify the impacts resulting from these proposals.

\vspace{0.3cm}
\noindent
{\bf Proposal 1: Divert the waters of the Limón stream to the Negro stream and install a sewage treatment plant at the mouth of the Negro stream.}

From the results above, we note that the elimination of faecal coliform discharges by the Limón and Negro streams would guarantee a pollution in Luruaco lake equal to that shown in Figure 3, when only the Mateo stream pollutes the lake. This scenario would guarantee that the water in the east sector of Luruaco lake, where the municipality of Luruaco is installed, would have very low concentrations of faecal coliforms. Therefore, our best proposal for the management of sewage discharged into Lake Luruaco would be to divert the waters from the Limón stream to the Negro stream, and install a sewage treatment plant at the mouth of the Negro stream. In this case, the pollution by faecal coliforms in Luruaco lake would be the one shown in Figure 3, with practically half of lake preserved from pollution by faecal coliforms.

In addition, this proposal is also a solution for the future sanitation of the municipality. Currently, there is an expansion process in the municipality of Luruaco, mainly towards the mouth of the Limón stream. This growth in the municipality will increase the discharge of sewage in the Limón stream and, consequently, the pollution in the Luruaco lake. In this scenario of expansion of the municipality, this proposal presents the best present and future solution for the management of sewage in Luruaco lake.

\vspace{0.3cm}
\noindent
{\bf Proposal 2: Divert the waters of the Limón and Negro streams to the Mateo stream.}

The second proposal is an alternative to the construction of a sewage treatment plant. This proposal would generate a scenario of pollution by faecal coliforms in Luruaco lake qualitatively similar to proposal 1. As in the first proposal, this scenario would also guarantee that the water in the east sector of Luruaco lake, where the municipality of Luruaco is installed, would have very low concentrations of faecal coliforms. 

\vspace{0.3cm}
\noindent
{\bf Proposal 3: Install a sewage treatment plant at the mouth of the Negro stream.}

The third proposal is based on the fact that the pollution injected by the Negro stream pollutes more the region of the mouth of the Limón stream than the reverse situation, that is, the pollution caused by the Limón stream in the mouth of the Negro stream, see Figures 4 and 5. In this context, our third proposal for sewage management is the installation of a single sewage treatment plant at the mouth of the Negro River.

\begin{figure}[H]
  	\begin{center}
  		\includegraphics[scale=0.8]{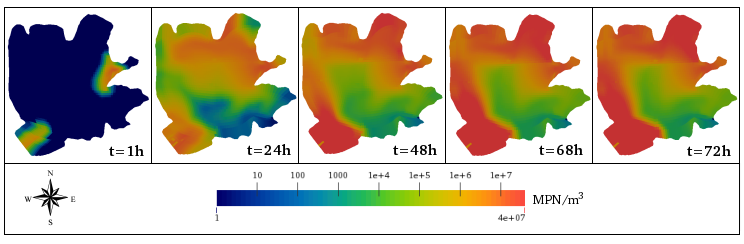} 
  		\caption{Proposal 3 for sewage management for Luruaco lake. Simulation of faecal coliform injection in Luruaco lake through Mateo and Limón streams, simultaneously. Font: Author.}
  		\label{fig7}
  	\end{center}
  \end{figure}

In this context, the following simulation considers the concentration of faecal coliforms in Mateo stream equal to $C_1=100 MPN/100 ml$, the concentration of faecal coliforms in Negro stream equal to $C_2=0 MPN/100 ml$, and the concentration of faecal coliforms in Limón stream equal to $C_3=500 MPN/100 ml$. Figure \ref{fig7} shows the faecal coliform concentrations in Luruaco lake due to our third proposal.
   
Comparing the simulations presented in figures 6 and 7, it is clearly observed the improvement of the water quality of Luruaco lake in the southeast and central regions of Luruaco lake.

Due to the characteristics of the hydrodynamic flow of the Luruaco Lake, we emphasize that the installation of single sewage treatment plant, in the other tributaries, does not generate significant improvements in the water quality of Luruaco Lake. 

\vspace{0.3cm}
\noindent
{\bf Proposal 4: Divert the waters of the Negro stream to the Mateo stream.}

The fourth proposal is also an alternative to the construction of a sewage treatment plant. It consists of diverting the Negro stream to the Mateo stream. The numerical simulations show that in this scenario the concentrations of faecal coliforms in Luruaco lake are qualitatively similar to those shown in Figure 7, and quantitatively slightly higher. Similarly to the third proposal, it also allows for an improvement in the water quality of the southeast sector of Luruaco lake. 

Due to the characteristics of the hydrodynamic flow of Lake Luruaco, we emphasize that the other diversion possibilities, from one stream to another stream, do not show local or global improvements in the water quality of Lake Luruaco.
  
\section{Conclusion}

In this study, the numerical simulations aimed to obtain a better understanding of the dynamics of faecal coliforms in Luruaco lake, and thus allow government agencies to create strategies to improve the water quality of Luruaco lake. We emphasize that all  proposals require studies of economic viability, technical viability, topographic viability, on environmental and ecological impacts, among other studies.

The results are also adjusted to the socioeconomic reality of the region. The municipality of Luruaco has only 3 $\%$ of residences served by a sewage network. Therefore, almost all of the wastewater in the municipality's residences is discharged in pits, which characterizes the low concentration of faecal coliforms in the Limon stream, $C_3=500 MPN/100 ml$. On the other hand, the massive use of these pits generates a great contamination in the groundwater.

Regarding the contamination of the water flow of the Negro and Mateo streams by faecal coliforms, it is essentially due to the feces of cattle and sheep. The regions surrounding these streams are areas of intensive agricultural and livestock activities, where there is no wastewater treatment. There is also no management of animal excrement or alternative use in the production of fertilizers. The concentrations of faecal coliforms in the Negro and Mateo streams are similar and in the order of $C_1=C_2=100 MPN/100 ml$.

Through the simulations performed in section 5, we were able to understand how some factors influence the dynamics of faecal coliforms in the lake, among which we highlight the influence of tributaries, the existence of hydrodynamic vortices and the conditions imposed on the model. Thus, the analysis of the results showed us which regions have the highest risk of contamination. In the current scenario, the southeast region of Lake Luruaco has the lowest concentrations of faecal coliforms. On the other hand, the lake region near the mouth of the Limón stream is the main source of water for residents of the municipality of Luruaco. However, the simulation shows that the concentration of faecal coliforms in this region is high, increasing the risk of contamination of the population. We suggest that water collection by the inhabitants of the municipality of Luruaco be carried out on the arm of the lake located between the Limón and Negro streams, where the numerical simulations show the lowest levels of contamination by faecal coliforms.

Following, we present proposals for the management of the sewage that is discharged into Lake Luruaco. The best sewage management presented in section 6 proposes that the Limón stream be diverted to the Negro stream and that a single sewage treatment plant be installed at the mouth of the Negro stream. In this scenario, numerical simulations show that the entire east region of Luruaco lake  would present very low contamination by faecal coliforms, as can be seen in Figure 3. 

An additional benefit of this proposal is to bring a solution for the future sanitation of the lake and the region. The objective of diverting the Limón stream to the Negro stream is to anticipate the expansion processes in the municipality of Luruaco that currently occur, mainly towards the mouth of the Limón stream. This expansion will increase the contamination and pollution of the lake. This proposal would be a solution to this future increase in contamination of Luruaco lake.

On the other hand, considering that this proposal may not be economically viable, we present three more proposals. 

The second proposal is an alternative to the first proposal. The second proposal is to divert the waters of the Limón and Negro streams to the Mateo stream. This proposal would generate a scenario of pollution by faecal coliforms in Luruaco lake qualitatively similar to the first proposal, with the entire east sector of the lake showing very low concentrations of faecal coliforms.

The third proposal is to install a single sewage treatment plant at the mouth of the Negro stream. The numerical simulations showed the best place to install a sewage treatment plant is at the mouth of the Negro stream. Comparing Figures 3 and 7, there is a big difference in water quality due to proposals 1 and 3. On the other hand, when comparing this proposal, Figure 7, with the current scenario, Figure 6, it appears that there is a certain improvement in water quality. This proposal, in addition to providing treated water to the municipality of Luruaco, coming from the treatment plant, it reduces the concentrations of fecal coliforms in the southeast sector of Lake Luruaco. Due to the characteristics of the hydrodynamic flow of Luruaco Lake, we emphasize that the installation of single sewage treatment plant, in the other tributaries, does not generate significant improvements in the water quality of Luruaco Lake. 
 
The fourth proposal is an alternative to the third proposal. The fourth proposal is to divert the waters of the Negro streams to the Mateo stream. The numerical simulations show that, in this scenario, the concentrations of faecal coliforms in Lake Luruaco are qualitatively similar to those presented in the third proposal, and quantitatively slightly higher. In this proposal, the region with the lowest contamination by faecal coliforms is the arm of the lake located between the Limón and Negro streams, a neighboring region of the municipality of Luruaco, which would facilitate the collection of water for the population. Due to the characteristics of the hydrodynamic flow of Luruaco Lake, we emphasize that the other diversion possibilities, from one stream to another stream, do not show local or global improvements in the water quality of Lake Luruaco.

Finally, we comment that the mathematical and computational model presented in this work can be applied to the other water bodies in the region.

\vspace {0.5cm}
\noindent
{\Large{\bf Acknowledgments}}
\vspace {0.5cm}

This project was supported by Universidade Estadual de Londrina (UEL) and Coordenação de Aperfeiçoamento de Pessoal de Nível Superior (CAPES). The author T. M. Saita thanks CAPES for the scholarship granted.

\vspace {0.5cm}
\noindent{\bf Conflicts of interest} - None
\vspace {0.5cm}


\vspace {0.5cm}
\noindent
{\Large{\bf{References}}}
\vspace {0.5cm}

\noindent
ASHBOLT, N.J., GRABOW, W.O.K., SNOZZI, M., 2001. Indicators of microbial water quality. In: Water Quality: Guidelines, Standards and Health. Risk assessment and management for water-related infectious disease, p.289-316, Eds. Fewtrell, L, and Bartram, J., IWA Press, London, Great Britain. 

\noindent
BRAUWERE, A., OUATTARA, N.K., SERVAIS, P., 2014. Modeling fecal indicator bacteria concentrations in natural surface waters: A review. Critical Reviews in Environmental Science and Technology, 44, 2380-2453. 

\noindent
BURDEN, R.L., FAIRES, J.D., BURDEN, A.M., 2015. Numerical Analysis, Cengage Learning, Boston, USA.

\noindent
BURNEO, P.C., GUNKEL, G., 2003. Ecology of a high Andean stream, Rio Itambi, Otavalo, Ecuador. Limnologica 33, 29-43.

\noindent
CHAPRA, S.C., 2008. Surface water-quality monitoring, Waveland Press, Inc, Long Grove, Illinois, United States.

\noindent
CIRILO, E.R., BARBA, A.N.D., NATTI, P.L., ROMEIRO, N.M.L., 2018. A numerical model based on the curvilinear coordinate system for the MAC method simplified. Semina: Exact and Technological Sciences, 39, 87-98.

\noindent
CRA - CORPORACIÓN AUTÓNOMA Y REGIONAL DEL ATLÁNTICO, 2007. Documentación del estado de las cuencas hidrográficas en el Departamento del Atlántico, Departamento del Atlántico, Colombia, CRA-Corporación Autónoma Regional del Atlántico,  Barranquilla, Colombia.

\noindent
CRA - CORPORACIÓN AUTÓNOMA Y REGIONAL DEL ATLÁNTICO, 2012. Diagnóstivo ambiental y estrategias de rehabilitación de  la Ciénega de Luruaco, Atlántico, Barranquilha, CRA-Corporación Autónoma Regional del Atlántico, Barranquilla, Colombia. 

\noindent
CRA - CORPORACIÓN AUTÓNOMA Y REGIONAL DEL ATLÁNTICO, 2014. Diagnóstico inicial para el ordenamiento del Embalse del Guájaro y la Ciénaga del Luruaco, Departamento del Atlántico, Colombia, CRA-Corporación Autónoma Regional del Atlántico,  Barranquilla, Colombia.

\noindent
DHSS - DEPARTMENT OF HEALTH AND SOCIAL SECURITY, 1982. The bacteriological examination of drinking water supplies. Methods for examination of waters and associated materials, 5th edition, Stationery Office Books, London, Great Britain.

\noindent
EVANS, G., BLACKLEDGE, J., YARDLEY, P., 1999. Numerical Methods for Partial Differential Equations, Springer-Verlag, London, Great Britain.

\noindent
FORTUNA, A.O., 2012. Técnicas computacionais para dinâmica dos fluidos: conceitos básicos e aplicações, EDUSP, São Paulo, Brazil.

\noindent
HELMER, R., HESPANHOL, I., 1997. Water pollution control: a guide to the use of water quality management principles, E \& FN Spon, London, Great Britain.

\noindent
HORITA, C.O., ROSMAN, P.C.C., 2006. A Lagrangian model for shallow water bodies contaminant transport. J. Coast. Res. 39, 1610–1613 (special issue).

\noindent
JANSSEN, P.H.M., HEUBERGER, P.S.C., 1995. Calibration of process-oriented models. Ecol. Model. 83, 55–66.

\noindent
JERVES-COBO, R., LOCK, K., BUTSEL, J.V., PAUTA, G., CISNEROS, F., NOPENS, I., GOETHALS, P.L.M., 2018. Biological impact assessment of sewage outfalls in the urbanized area of the Cuenca River basin (Ecuador) in two different seasons. Limnologica 71, 8-28.

\noindent
LIU, W.C., CHAN, W.T., YOUNG, C.C., 2015. Modeling fecal coliform contamination in a tidal Danshuei River estuarine system.
Science of The Total Environment, 502, 632-640.

\noindent
MALISKA, C.R., RAITHBY, G.D., 1984. A method for computing three dimensional flows using non‐orthogonal boundary‐fitted co‐ordinates. 
International Journal for Numerical Methods in Fluids, 4, 519-537.

\noindent
MALISKA, C.R., 2004. Transferência de calor e mecânica dos fluidos computacional, LTC, Rio de Janeiro, Brazil.

\noindent
MCKEE, S., TOME, M.F., FERREIRA, V.G., CUMINATO, J.A., CASTELO, A., SOUSA, F.S., MANGIAVACCHI, N., 2008. The MAC method. Computers and Fluids, 37, 907-930.

\noindent
MUNICIPIO DE LURUACO, 2012. Plan de Desarrollo Municipal de Luruaco 2012-2015, Luruaco, Departamento del Atlántico, Colombia.

\noindent
OPS - ORGANIZACION PANAMERICAN DE LA SALUD, 1988. Guías para la calidad del agua potable, thirty edition, volume 3, Editor de la Organizacion Panamericana de la Salud, Washington, D.C, United State.

\noindent
OUATTARA, N.K., YAO, C.K., KAMAGATÉ, B., DROH, L.G., OUATTARA, A., GOURÈNE, G., 2018. Impact of faecal bacteria contamination on drinking
water supply in Aghien Lagoon, Abidjan, Ivory Coast. Afr. J. Microbiol. Res.,12 , 965-972.

\noindent
PARDO, S.R., NATTI, P.L., ROMEIRO, N.M.L., CIRILO, E.R., 2012. A transport modeling of the carbon-nitrogen cycle at Igapó I Lake - Londrina, Paraná State, Brazil. Acta Scientiarum. Technology, 34, 217-226.

\noindent
RECKHOW, K.H., CHAPRA, S.C., 1983. Confirmation of water quality models. Ecol. Model. 20, 113–133.

\noindent
ROHATGI, A., 2018. WEbPlotDigitizer Version 4.1, Austin, Texas, United States.

\noindent
ROMEIRO, N.M.L., CASTRO, R.G.S., CIRILO, E.R., NATTI, P.L., 2011. Local calibration of coliform parameters of water quality problem at Igapó I Lake - Londrina, Paraná, Brazil. Ecological Modelling, 222, 1888-1896.

\noindent
ROMEIRO, N.M.L., MANGILI, F.B., COSTANZI, R.N., CIRILO, E.R., NATTI, P.L., 2017. Numerical simulation of BOD5 dynamics in Igapó I lake, Londrina, Paraná, Brazil: Experimental measurement and mathematical modeling. Semina: Exact and Technological Sciences, 38, 50-58.

\noindent
SAITA, T.M., NATTI, P.L., CIRILO, E.R., ROMEIRO, N.M.L., CANDEZANO, M.A.C., ACUNA, R.A.B., MORENO, L.C.G., 2017. Simulação numérica da dinâmica de coliformes fecais no lago Luruaco, Colômbia. Tendências em Matemática Aplicada e Computacional, 18, 435-447.

\noindent
SHAKIB, F.; HUGHES, T. J. R.; JOHAN, Z., 1991. A new finite element formulation for computational fluid dynamics: X. The compressible Euler and Navier-Stokes equations. Computer Methods in Applied Mechanics and Engineering, 89, 141-219.

\noindent
VISHNURADHAN, R., ELDHO, T.I., VETHAMONY, P., SAHEED, P.P., SHIRODKAR, P.V., 2018. Assessment of the environmental health of an ecologically sensitive, semi-enclosed, basin - A water quality modelling approach. Marine Pollution Bulletin, 137, 418-429.

\noindent
WANG, M., XU, X., WU, Z., ZHANG X., SUN, P., WEN, Y., WANG, Z., LU, X., ZHANG, W., WANG, X., TONG, Y., 2019. 
Seasonal Pattern of Nutrient Limitation in a Eutrophic Lake and Quantitative Analysis of the Impacts from Internal Nutrient Cycling. Environmental Science and Technology, 53, 13675-13686.

\noindent
WHO - World Health Organization, 2017. Guidelines for drinking-water quality, fourth edition, WHO Press, Geneva, Switzerland.
							
\end{document}